\documentclass[10pt, conference, compsocconf]{IEEEtran}
\IEEEoverridecommandlockouts
\usepackage{cite}
\usepackage{amsmath,amssymb,amsfonts}
\usepackage{algorithmic}
\usepackage{graphicx}
\usepackage{textcomp}
\usepackage{xcolor}
\usepackage{graphicx}
\usepackage{caption}
\usepackage{subcaption}
\usepackage{hyperref}
\usepackage{amsmath}
\usepackage{accents}
\usepackage{multicol}
\usepackage{balance} 
\usepackage{subcaption} 
\usepackage{newunicodechar}
\usepackage{adjustbox}
\usepackage{booktabs}
\usepackage{multirow}
\usepackage{graphicx}
\usepackage{xspace}
\usepackage{amssymb}

\newunicodechar{−}{-}     
\def\BibTeX{{\rm B\kern-.05em{\sc i\kern-.025em b}\kern-.08em
    T\kern-.1667em\lower.7ex\hbox{E}\kern-.125emX}}

\newcommand{\Algoname}{KiNETGAN\xspace}
    
\begin{document}

\title{\Algoname: Enabling Distributed Network Intrusion Detection through Knowledge-Infused Synthetic Data Generation}

\author{\IEEEauthorblockN{ Anantaa Kotal}
\IEEEauthorblockA{\textit{C.S.E.E.} \\
\textit{University of Maryland,} \\
\textit{Baltimore County}\\
Baltimore, USA \\
anantak1@umbc.edu}
\and
\IEEEauthorblockN{ Brandon Luton}
\IEEEauthorblockA{\textit{C.S.E.E.} \\
\textit{University of Maryland,} \\
\textit{Baltimore County}\\
Baltimore, USA \\
cl39497@umbc.edu}
\and
\IEEEauthorblockN{ Anupam Joshi}
\IEEEauthorblockA{\textit{C.S.E.E.} \\
\textit{University of Maryland,} \\
\textit{Baltimore County}\\
Baltimore, USA \\
joshi@umbc.edu}
}
\maketitle

\begin{abstract}
In the realm of IoT/CPS systems connected over mobile networks, traditional intrusion detection methods analyze network traffic across multiple devices using anomaly detection techniques to flag potential security threats. However, these methods face significant privacy challenges, particularly with deep packet inspection and network communication analysis. This type of monitoring is highly intrusive, as it involves examining the content of data packets, which can include personal and sensitive information. Such data scrutiny is often governed by stringent laws and regulations, especially in environments like smart homes where data privacy is paramount. Synthetic data offers a promising solution by mimicking real network behavior without revealing sensitive details. Generative models such as Generative Adversarial Networks (GANs) can produce synthetic data, but they often struggle to generate realistic data in specialized domains like network activity. This limitation stems from insufficient training data, which impedes the model's ability to grasp the domain's rules and constraints adequately. Moreover, the scarcity of training data exacerbates the problem of class imbalance in intrusion detection methods. To address these challenges, we propose a Privacy-Driven framework that utilizes a knowledge-infused Generative Adversarial Network for generating synthetic network activity data (\Algoname). This approach enhances the resilience of distributed intrusion detection while addressing privacy concerns. Our Knowledge Guided GAN produces realistic representations of network activity, validated through rigorous experimentation. We demonstrate that \Algoname maintains minimal accuracy loss in downstream tasks, effectively balancing data privacy and utility.

\end{abstract}


\section{Introduction}
Network intrusion detection systems (NIDS) are critical for protecting modern enterprise systems, particularly in IoT-based and mobile environments where attacks can lead to both information loss and physical damage. Distributed NIDS enable real-time monitoring across multiple devices and segments, promptly detecting anomalies and potential threats to safeguard sensitive data. Integrating Machine Learning (ML) models in NIDS enhances their effectiveness in preventing cyberattacks \cite{da2019internet, ucci2019survey, ding2018survey, piplai2020creating, piplai2020using, dasgupta2020comparative, das2023change}.

However, sharing data across distributed systems raises privacy concerns, especially with intrusive detection methods like deep packet inspection. Federated learning offers a solution by allowing collaborative training without sharing raw data, but it's challenging to implement across diverse devices with varying architectures and learning infrastructure.

Deep learning-based synthetic data generation emerges as a promising solution. By creating data that mirrors authentic network behavior while protecting sensitive details, it enables secure data sharing and analysis. Among these methods, generative adversarial networks (GANs) stand out for their ability to capture and replicate the underlying distribution of a training dataset.

However, standard GANs face limitations in generating realistic system data, particularly for IoT and mobile networks, due to a lack of explicit domain knowledge. For instance, they may misconfigure attributes like port numbers associated with specific attacks, leading to misleading synthetic data. Moreover, class imbalance in training data can bias models towards prevalent classes, hindering accurate intrusion detection.

To address these challenges, leveraging domain knowledge to guide generative models is crucial. By incorporating specific characteristics of the data into the training process, such as rules governing network traffic, generative models can produce more accurate synthetic data. This approach enhances the efficacy of generative models in specialized domains like network security by ensuring that synthetic data closely resembles real-world scenarios.

This paper introduces a novel Privacy-Driven knowledge-infused Generative Adversarial Network (\Algoname) model designed to tackle the obstacles related to synthetic data generation for privacy preservation in distributed network intrusion detection systems. Our innovative approach leverages domain knowledge and employs enhanced GAN training to create realistic representations of network activities on individual devices. The \Algoname model addresses the limitations of standard generative models, ensuring a comprehensive understanding of the domain's rules and restrictions. We demonstrate the effectiveness of our approach by synthesizing network activity data, validating the synthetic dataset against network-specific constraints, and confirming its suitability through likelihood fitness and high efficacy in downstream intrusion detection tasks.

\section{Background}
\label{sec_background}
Generative Adversarial Networks (GANs) are powerful models widely used for generating synthetic data that closely resembles real data \cite{brock2018large, isola2017image, zhang2017stackgan, wang2018high}. They consist of a generative model (G) and a discriminative model (D) trained together in a min-max game framework. GANs have demonstrated high accuracy in generating synthetic data, particularly for images and text. However, teaching GANs to learn from Network Activity Data presents challenges due to its tabular nature, combining discrete and continuous values, and exhibiting sparsity and imbalanced distributions.

Xu et al. proposed a GAN model addressing challenges with tabular data by introducing mode-specific normalization, a conditional generator, and training by sampling \cite{xu2019modeling}. Kotal et al. extended this model for privacy-preserving data generation, enforcing t-closeness in the synthetic data distribution to preserve privacy \cite{kotal2022privetab}.

Differential Privacy (DP) has been applied to GANs to enhance privacy \cite{abadi2016deep}. Various models combine DP with GANs to generate differentially private synthetic data, introducing noise into the discriminator during training to ensure privacy \cite{xie2018differentially, torkzadehmahani2019dp}.

However, Network traffic data poses challenges for GAN training due to sparsity and limited size. GANs trained solely on observed network activity lack understanding of network attributes and struggle to adhere to strict rules without explicit constraints. Knowledge guidance becomes essential to convey constraints and enhance the data generation process.

Knowledge Graphs (KGs) offer a versatile data model for knowledge representation and reasoning \cite{hui2022knowledge}. They store contextual information crucial for learning in distributed systems and can impose constraints on entities. Integrating KGs enriches the data generation process, improving contextual awareness in ML systems. Its integration has been demonstrated to significantly improve contextual awareness in machine learning (ML) systems \cite{piplai2020creating, narayanan2018early, elluri2021policy, piplai2023knowledge}. In a related context, Hui et al. introduced a knowledge-enhanced GAN for generating IoT traffic data \cite{hui2022knowledge}. Specifically, for privacy preservation, there is evidence that Knowledge Infusion can help generative models gain the added context needed for zero-shot learning and learning with limited parameters \cite{kampffmeyer2019rethinking, chen2021knowledge, kotal2023knowledge, kotal2023privacy}.

In this paper, knowledge about network traffic is injected into GAN training by adding the Knowledge base as an independent discriminator. This approach enhances the GAN's understanding of network attributes and adherence to protocol rules, improving the accuracy of synthetic data generation in the observed system.

\begin{figure}
        \includegraphics[width=\columnwidth]{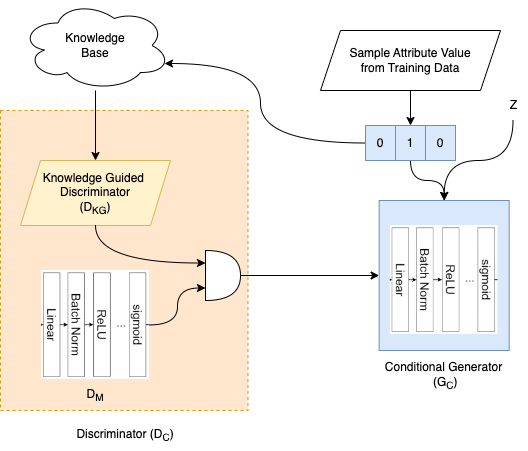} 
        \caption{The \Algoname model of Knowledge Infused Synthetic Data Generation}
        \label{fig_overall_arch}
\end{figure}

\section{Proposed Framework}
\label{sec_proposed_framework}
Network traffic data presents challenges for synthetic data generation due to sparsity, class imbalance, and strict domain rules. These characteristics hinder accurate model construction solely based on observed data, leading to unrealistic synthetic data. Domain knowledge integration into generative models reduces the need for relearning rules, addressing class and domain restrictions. We propose a knowledge-guided synthetic data generation method, \Algoname, leveraging a domain Knowledge Graph (KG) to train Generative Adversarial Networks (GANs) and conditional GANs. This approach tackles class imbalance and attribute cross-correlation issues. The framework's architecture is detailed in Figure \ref{fig_overall_arch}.

\subsection{Conditional Generator with Data Balancing}
In GANs, training on randomly selected data can lead to underrepresentation of minority categories, affecting generator accuracy. Conditional generation is vital for Mobile Network Activity data, enabling the GAN to learn attribute relationships and address data imbalance. Efficient data resampling ensures balanced category representation during training, maintaining original data distribution fidelity. We propose Conditional GANs and sampling-based training to achieve these goals, preserving model fidelity and accurately reflecting the original data distribution during testing.

\subsubsection{Conditional vector} To introduce the Conditional Attributes as the condition to the Conditional generator, we introduce the condition vector, \textbf{\textit{C}}. It is a one hot vector representation of the set of discerete Attributes. Let  $(c_1, c_2, ..., c_n)$ be the list of conditional attributes and  is the output from $G_C$ that is the condition for our current generation. 

Let us consider the attribute $c_1$. Let the range for $c_1$ be $\{c_{1,1}, c_{1,2}, ..., c_{1,n}\}$ The chosen value for $c_1$ is $\hat{c_1}$. The one hot vector representation ($C_1$) of $c_1$ is defined as follows: 

\begin{equation}
        C_{1,i} = 
        \begin{cases}
            1,  & \text{if } c_{1,i} == \hat{c_1}\\
            0,  & \text{otherwise}
        \end{cases}
\end{equation}

The conditional vector, \textbf{\textit{C}} is a concatenation of all $C_i$'s:

\begin{equation}
    C = C_1 \oplus	C_2 \oplus	... \oplus	 C_n
\end{equation}

\subsubsection{Conditional Generator}
The input to the Generator is a random noise (Z) and the conditional vector $\textbf{\textit{C}}$. The objective of the Conditional Generator is to generate realistic synthetic data while adhering to the attribute values specified in  $\textbf{\textit{C}}$. To ensure that this constrained is met, in addition to the discriminator score, we need to penalise the generator for disregarding the attribute values specified in $\textbf{\textit{C}}$. This is done by adding a cross entropy with the condition vector, $\textbf{\textit{C}}$ to the loss function. Let the generator output for the conditional attribute set be $\hat{c} = (\hat{c_1}, \hat{c_2}, ..., \hat{c_n})$, the one hot vector representation of which is \textbf{\textit{$\hat{C}$}}. Then we add the following term to the loss function of Conditional Generator: \textit{BCE}$($\textbf{\textit{C}} , \textbf{\textit{$\hat{C}$}}$)$ averaged over all the instances of the batch. As the training advances, the generator learns to make an exact copy of \textbf{\textit{C}} into \textbf{\textit{$\hat{C}$}}.

\subsubsection{Conditioning on Imbalanced Values} As discussed, mobile network activity data is often sparse, and attribute values are heavily imbalanced. To ensure that minority attribute values are sufficiently represented during training, we need to compel the generator to consider these minority values. To achieve this, we randomly sample an attribute value from a uniform distribution within the range of the attribute and add it to our condition vector, \textbf{\textit{C}}. The generator is thus constrained to produce synthetic data where the minority attribute is present. This ensures that sparse attribute values are adequately represented during the training of the generator, enabling it to produce data points within these values.

\subsection{GAN Training with Knowledge-Guided Discriminator}
\label{subsec_KGGAN}
The Knowledge-Guided Discriminator aims to identify external rules restricting attribute values using a Knowledge Graph. It focuses on attributes with relative value restrictions to learn valid and invalid attribute combinations. For example, in the CVE-1999-0003 attack, a valid port address is between 32771 and 34000. Domain knowledge helps rule out invalid combinations, such as ports outside this range. The key difference is that some generator outputs may not just be fake but invalid. Penalizing these instances ensures the generator produces realistic and valid instances. This objective is met by dividing the discriminator into two parts.

\begin{enumerate}
    \item \textbf{Knowledge-Guided Discriminator (D\textsubscript{KG}):} The objective for the Knowledge-Guided Discriminator is to discriminate between correct and incorrect instances of data according to domain rules. In the case of network activity data, this includes examples of invalid combinations of IP addresses and port numbers for an event. A domain-specific Knowledge Graph (KG) can help rule out invalid combinations of attributes from explicit knowledge. The KG is queried with the output ($\hat{c_1}, \hat{c_2}, ..., \hat{c_n}$) of $G_C$ to determine whether the given set of values is valid. Let us represent the KG query as Q. The discriminator's input consists of all valid sets of attributes for the conditional vector \textbf{\textit{C}} queried from the knowledge graph and the output of the generator, $G_C$.
    \item  \textbf{Regular Discriminator:} The objective here is that of a regular discriminator in a GAN model. It is a standard discriminator tasked with distinguishing real data points from those generated by the generator, $G_C$. This framework follows the design of a standard GAN. The input to the discriminator $D_M$ includes the output of $G_C$ and real data points from the training set.
\end{enumerate}

The final output of the discriminator ($D_C$) combines the output of both $D_{KG}$ and $D_M$:
\begin{equation}
    D_C = D_{KG} + D_M
\label{combined_D_value}
\end{equation}

\textbf{Loss function:}
The generator loss is updated on the output of both $D_{KG}$ and $D_M$. Thus following from equation \ref{combined_D_value}, the loss for $G_C$ is defined as:
\begin{equation}
    \mathcal{L}_{G_C} = E_{z \sim p_z(z)}[log(1 - D_C(G_C(z)))]
\end{equation}

\begin{figure}
        \includegraphics[width=\columnwidth]{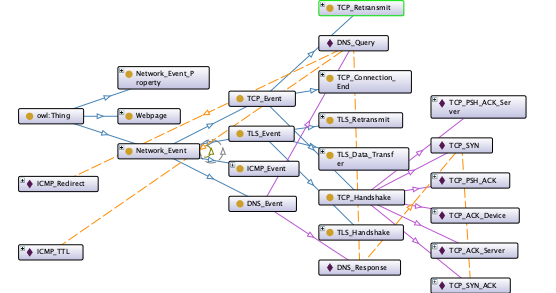} 
        \caption{Ontology for Network Activity Capture}
        \label{fig_network_kg}
\end{figure}

\section{Experimental Framework}
\label{sec_exp_method}
\subsection{Knowledge Graph Creation}
The Unified Cybersecurity Ontology (UCO) is a comprehensive framework designed to enhance cyber situational awareness by integrating diverse data and knowledge schemas from various cybersecurity systems and standards. UCO encompasses entities, events, activities, and relationships crucial for cybersecurity analysis. Integrating UCO into machine learning models improves contextual understanding, enhancing their effectiveness in cybersecurity scenarios. This study extends UCO to define concepts in network activity data, introducing entities like "networkEvent" and "domainURL". Each network event is defined by properties such as protocol, source/destination IP addresses, and port numbers. Leveraging this ontology, a Network Traffic Knowledge Graph (NetworkKG) is constructed, guiding data generation. A Knowledge Graph reasoner facilitates queries for valid IP, port, and protocol combinations, assisting the Generative Adversarial Network (GAN) synthesis process. This approach ensures that synthetic data generated aligns with real-world network event attributes, enhancing the quality and relevance of generated data for cybersecurity applications. Figure \ref{fig_network_kg} demonstrates the entities in this ontology. 

\subsection{Data Collection}
\label{subsec_data}
To test our method, we require datasets of real network activity from a system of devices. For this purpose, we utilize two datasets: a network activity dataset collected from a system of IoT devices connected in our lab and the UNSW-NB15 Dataset.

\subsubsection{Lab Collected Data:}
In our network setup, we've integrated mobile and IoT devices such as a Blink camera, a smart plug, and a motion sensor. Analyzing their communication patterns with Wireshark, we focus on events like motion detection, lamp activation, and tag manager interactions. Collected data includes Source/Destination IP, Ports, and Protocols. Filtering by device IPs, we study their typical communications and simulate attacks like Traffic flooding. The dataset, comprising 14,520 records, is crucial for NIDS training.

\subsubsection{UNSW-NB15 Data:}
The UNSW-NB15 dataset consists of 2,540,044 network connection records. This comprehensive dataset includes a wide range of network traffic data, featuring 49 attributes that encompass flow features, basic features, content features, time features, and additional generated features. This size and diversity make it well-suited for training and evaluating machine learning algorithms for intrusion detection systems.

\section{Experimental Results}
\label{sec_exp_results}
In this section, we will begin by describing and detailing the various measures and tests we used to validate our model. The main goal of network activity datasets is to support Network Intrusion Detection (NIDS) efforts. Machine learning-based NIDS classifiers require high fidelity data for training. As previously mentioned, the bottleneck for security efforts lies in the challenge of obtaining dependable training data. To fulfill its purpose, synthetic data must serve as a viable substitute for the original data in downstream tasks.

To demonstrate that the KiNETGAN framework fulfills these objectives, we present the outcomes of three types of tests:
\begin{itemize}
    \item \textbf{Fidelity Results} to show that the synthetic data is statistically close or similar to the original data.
    \item \textbf{Utility Results} to show that the synthetic data is useful in training downstream ML-based NIDS models.
    \item \textbf{Privacy Results} to show that the synthetic data and \Algoname model are resilient against privacy attacks.
\end{itemize}

We demonstrate that the \Algoname framework fulfills these objectives by being distributionally similar to the original dataset and having comparable accuracy in downstream tasks. To validate our model, we compare it with synthetic data generated using other Generative Deep Learning models for tabular data generation: CTGAN \cite{xu2019modeling}, OCTGAN \cite{kim2021oct}, PATEGAN \cite{jordon2018pate}, TABLEGAN \cite{park2018data}, and TVAE \cite{xu2019modeling}. We provide experimental results for synthetic data generated from training on our dataset of lab-collected network traffic data and the UNSW-NB15 data.

\begin{table}[]
\centering
\resizebox{0.9\columnwidth}{!}{%
\begin{tabular}{l|ll|ll|}
\cline{2-5}
                                     & \multicolumn{2}{c|}{Lab Data}                      & \multicolumn{2}{c|}{UNSW-NB15}                     \\ \cline{2-5} 
                                     & \multicolumn{1}{l|}{EMD}           & Distance      & \multicolumn{1}{l|}{EMD}           & Distance      \\ \hline
\multicolumn{1}{|l|}{CTGAN}          & \multicolumn{1}{l|}{{0.06}} & 0.09          & \multicolumn{1}{l|}{{0.07}} & 0.2           \\ \hline
\multicolumn{1}{|l|}{OCTGAN}         & \multicolumn{1}{l|}{1.61}          & 0.95          & \multicolumn{1}{l|}{1.32}          & 1.61          \\ \hline
\multicolumn{1}{|l|}{PATEGAN}        & \multicolumn{1}{l|}{1.07}          & 0.09          & \multicolumn{1}{l|}{0.53}          & 0.24          \\ \hline
\multicolumn{1}{|l|}{TABLEGAN}       & \multicolumn{1}{l|}{1.02}          & 0.19          & \multicolumn{1}{l|}{1.21}          & 0.53          \\ \hline
\multicolumn{1}{|l|}{TVAE}           & \multicolumn{1}{l|}{0.06}          & {0.04} & \multicolumn{1}{l|}{0.13}          & {0.23} \\ \hline
\multicolumn{1}{|l|}{\textbf{\Algoname}} & \multicolumn{1}{l|}{\textbf{0.06}}  & \textbf{0.03} & \multicolumn{1}{l|}{\textbf{0.07}} & \textbf{0.03} \\ \hline
\end{tabular}%
}
\caption{Comparison of Distance between Synthetic and Original Data}
\label{tab:distance}
\end{table}

\begin{figure} [t]
    \centering
    \includegraphics[width=0.9\columnwidth]
    {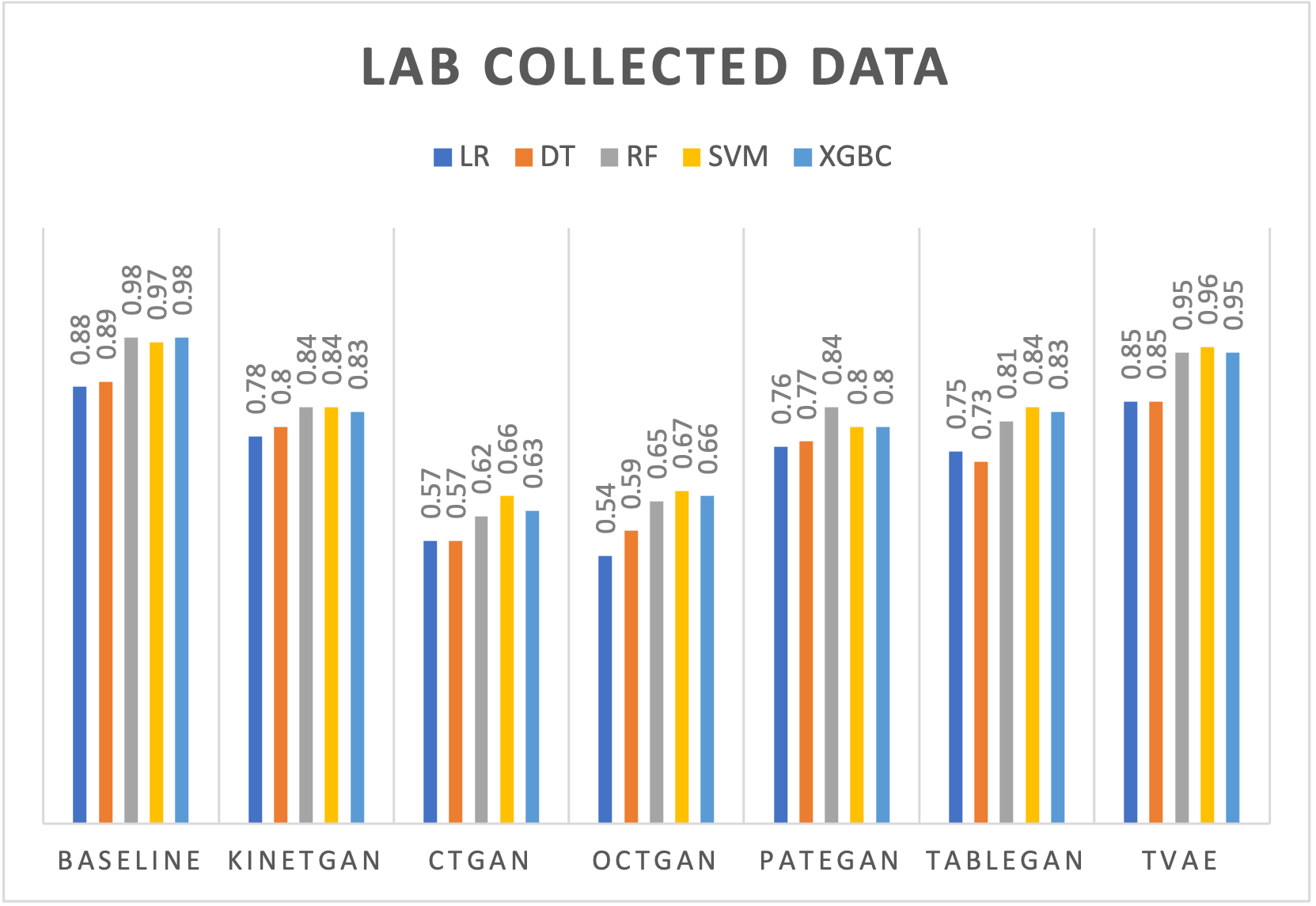}
    \caption{Comparison of NIDS accuracy for Lab Collected Data}
    \label{fig:accuracy_lab}
\end{figure}

\begin{figure} [t]
    \centering
    \includegraphics[width=0.9\columnwidth]
    {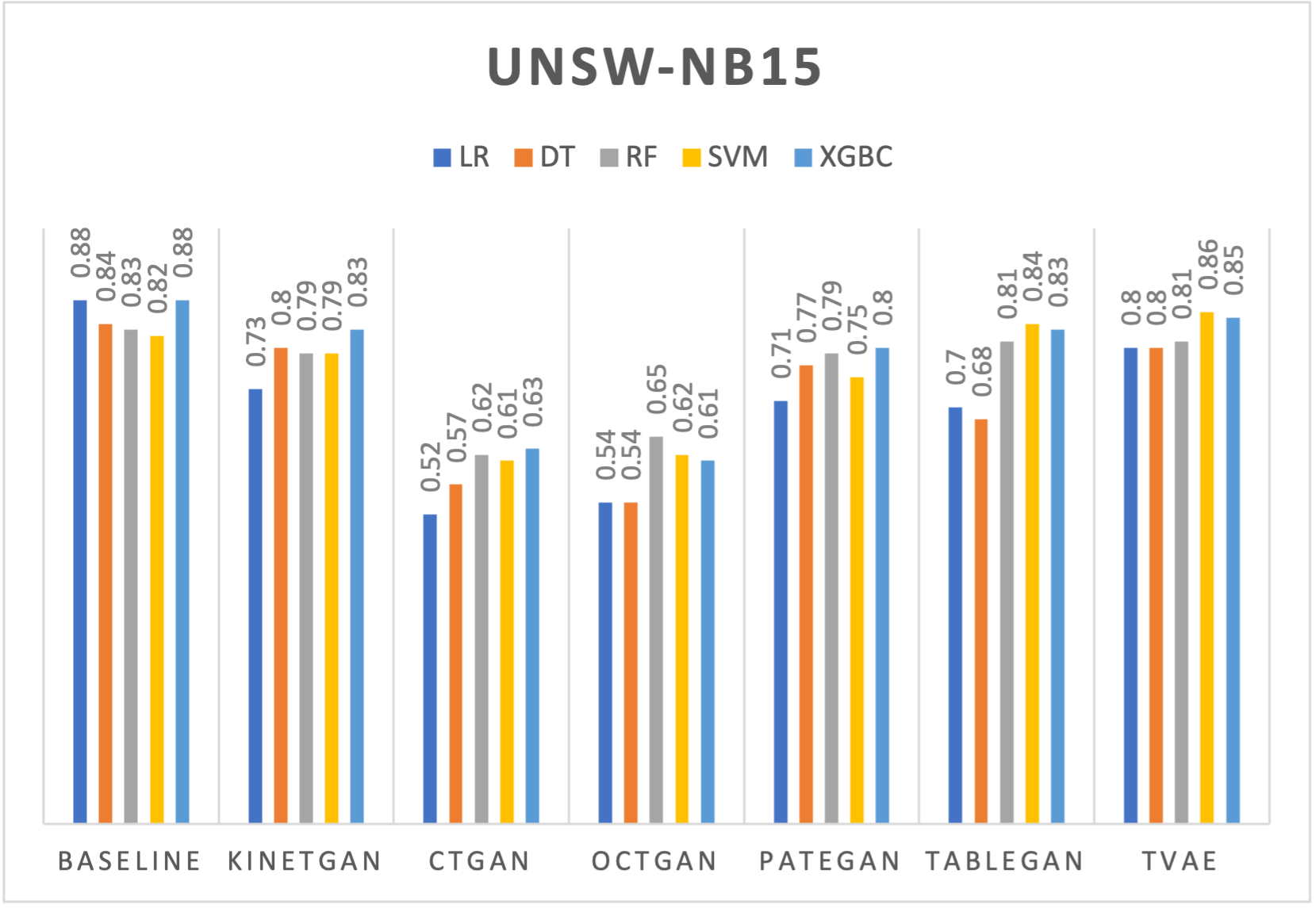}
    \caption{Comparison of NIDS accuracy for UNSW-NB15}
    \label{fig:accuracy_unsw}
\end{figure}

\subsection{Statistical Distance}
When assessing the quality of generated data, statistical distance measures help quantify the dissimilarity between the generated data distribution and the original (real) data distribution. We use two distance metrics for our comparison. 
\begin{itemize}
    \item The Earth Mover's Distance (EMD) or Wasserstein Distance which measures the minimum cost of turning one distribution into another, where the cost is interpreted as the amount of "mass" that must be moved.
    \item We use a combination of $L_1$ norm or  Manhattan distance to calculate the distance for categorical variables and the $L_2$ norm or  Euclidean distance to calculate the distance for continuous variables. Since our data is tabular i.e. a mix of categorical and continuous variables, this pragmatic approach is ideal to handle mixed-type data. 
 \end{itemize}

 The results from our comparison are given in Table \ref{tab:distance}. For the Lab collected data, the \Algoname had the lowest EMD distance at 0.06, similar to TVAE and CTGAN. \Algoname had the lowest combined distance at 0.032.  For UNSW-NB15 \Algoname and CTGAN had the lowest EMD distance at 0.007. and the \Algoname model has a combined distance of 0.16.

\subsection{Utility Results}
Machine Learning (ML) is essential in network intrusion detection, identifying anomalies indicating security threats. Network activity datasets aid ML models by providing extensive training data. We evaluate synthetic data's efficacy in replacing original data by training ML classifiers on both. \Algoname, our proposed synthetic data generation model, demonstrates competitive accuracy. Figure \ref{fig:accuracy_lab} demonstrates the accuracy for the baseline classifier with the classifiers trained on synthetic data from generated models including the \Algoname model on lab collected data. On lab collected data, \Algoname achieves an average accuracy of 0.81, surpassing other models like CTGAN and TableGAN. Figure \ref{fig:accuracy_unsw} demonstrates the accuracy for the baseline classifier with the classifiers trained on synthetic data from generated models including the \Algoname model on UNSW-NB15 data. For the UNSW-NB15 dataset, \Algoname achieves an average accuracy of 0.78, outperforming competing models. These results confirm \Algoname's potential in generating synthetic data for NIDS applications. Notably, \Algoname's accuracy surpasses other tabular data models like CTGAN and OCTGAN. This suggests that \Algoname can effectively replace original data in downstream tasks, enhancing the robustness of NIDS classifiers.

\subsection{Privacy Results}
Re-identification, Attribute Inference, and Membership Inference attacks are significant privacy threats targeting machine learning models. In our experiments, we observe the effectiveness of KiNETGAN in mitigating these risks. 

\begin{figure} [t]
    \centering
    \includegraphics[width=0.9\columnwidth]
    {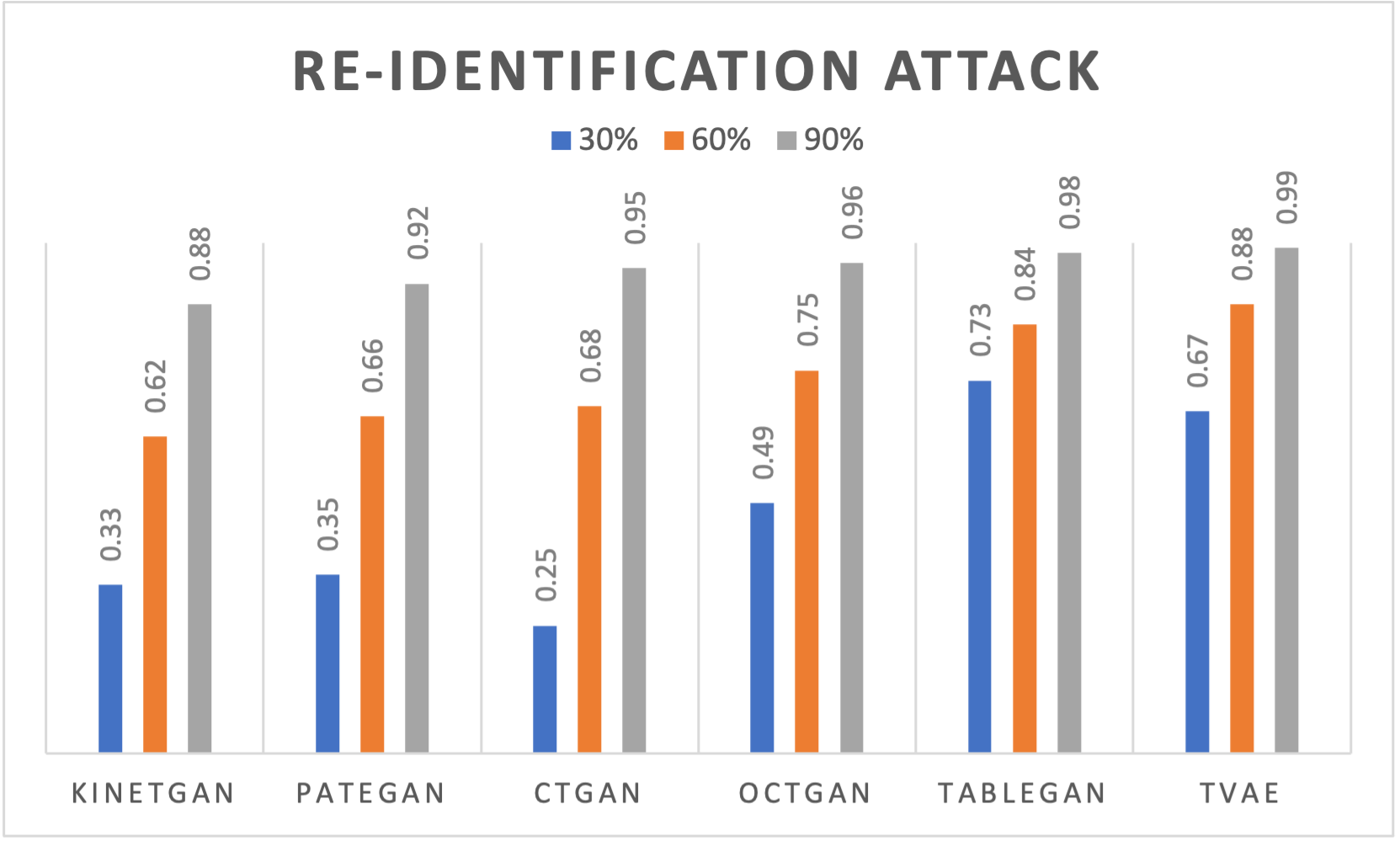}
    \caption{Comparison of Re-identification Attack with 30\%, 60\% and 90\% overlap on original data}
    \label{fig:reident}
\end{figure}

The Re-identification attack aims to link de-identified data with additional knowledge to reveal sensitive attributes. In Figure \ref{fig:reident}, we present the results for the accuracy with which the attack model is able to uniquely identify data points assuming it has prior knowledge about 30\%, 60\% and 90\% of the lab collected data. KiNETGAN outperforms other models, achieving an attack accuracy of 0.62 and 0.88 with 60\% and 90\% overlap with the data respectively. 

\begin{figure} [t]
    \centering
    \includegraphics[width=0.9\columnwidth]
    {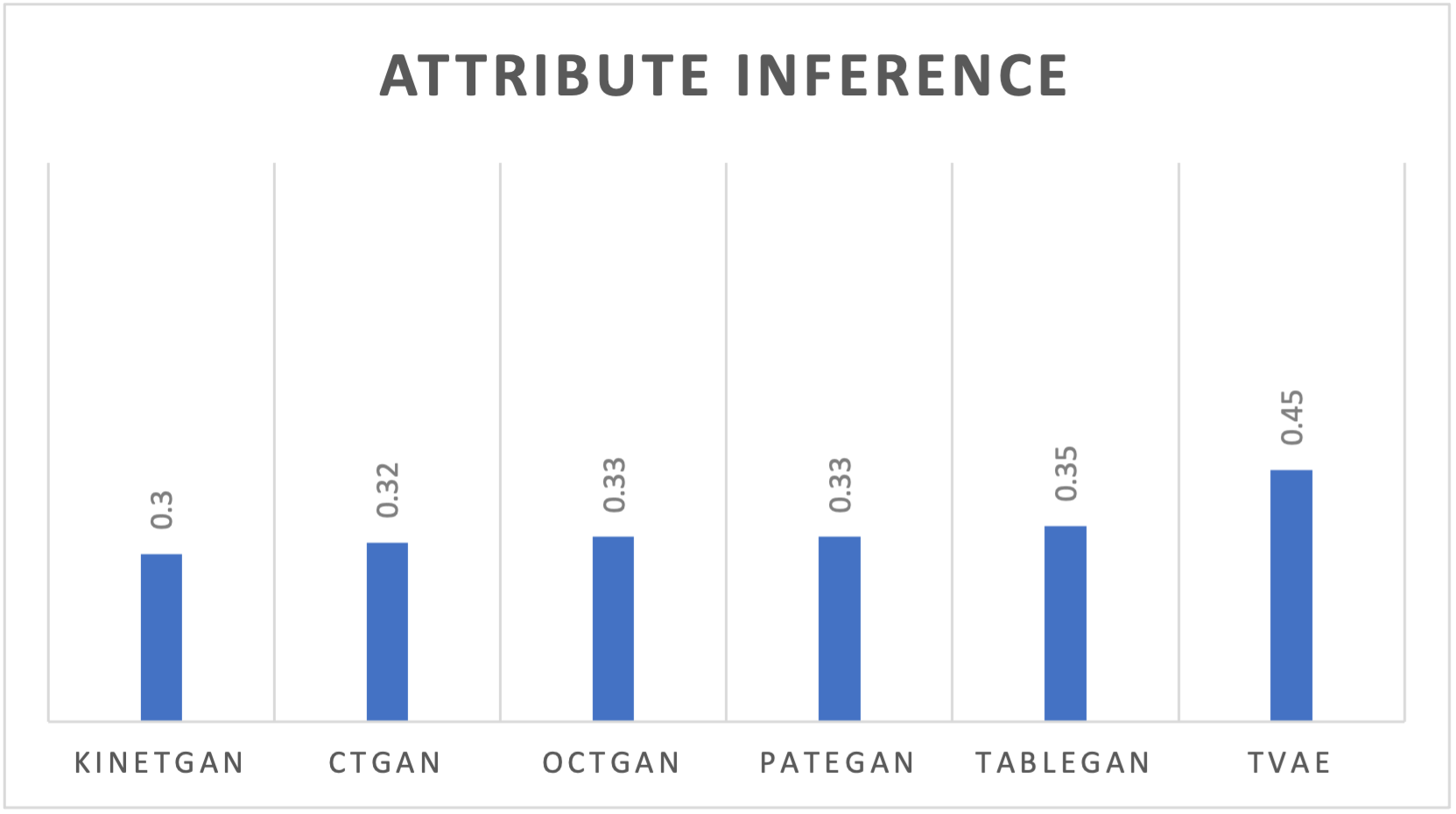}
    \caption{Comparison of accuracy in Attribute Inference Attack}
    \label{fig:attrib_infer}
\end{figure}

Attribute Inference attacks deduce sensitive attributes by analyzing seemingly innocuous data points. Figure \ref{fig:attrib_infer} shows the results for the Attribute Inference attack on synthetic data for the lab collected dataset.  KiNETGAN exhibits resilience, with an attack accuracy of 0.3.

\begin{figure} [t]
    \centering
    \includegraphics[width=0.9\columnwidth]
    {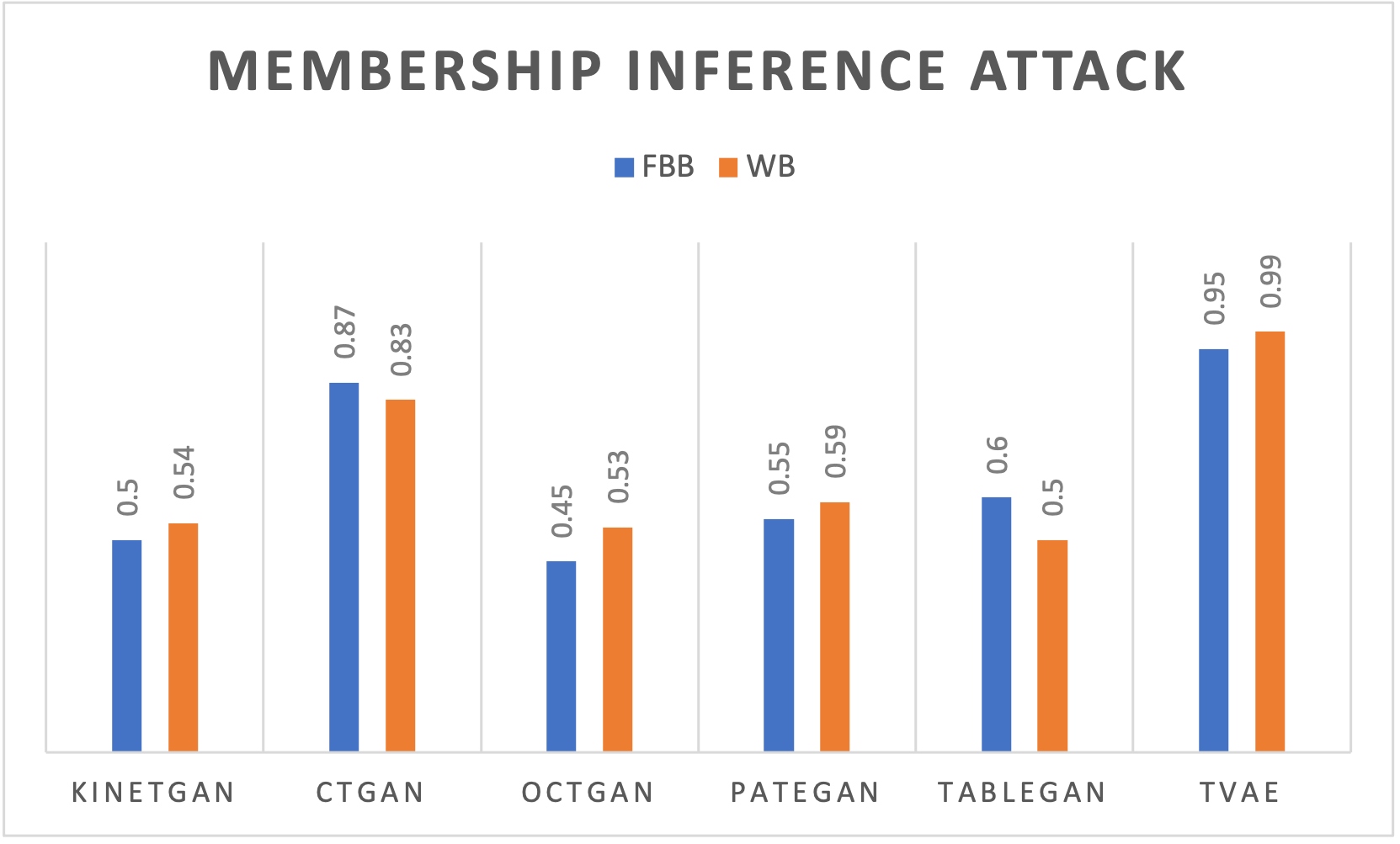}
    \caption{Comparison of Membership Inference Attack in White Box (WB) and FBB (Fully Black Box) setting}
    \label{fig:mia}
\end{figure}

Membership Inference attacks determine if a specific data point was part of the model's training dataset. Figure \ref{fig:mia} shows the results of membership inference attack in WB and FBB setting against lab collected data.  KiNETGAN demonstrates resilience in both White-Box (0.54) and Fully Black Box (0.5) settings, outperforming other models like CTGAN and TableGAN. These findings underscore KiNETGAN's effectiveness in preserving privacy and security in synthetic data generation.

\section{Conclusion}
\label{sec_conclusion}
To foster collaboration and address privacy concerns, this paper explores a novel knowledge-infused Generative Adversarial Network model for network activity data (\Algoname). Leveraging domain knowledge and enhanced GAN training, \Algoname overcomes challenges of domain restriction, class imbalance, and privacy preservation, creating realistic representations of network activities. We demonstrate the efficacy of \Algoname through synthetic dataset validation and likelihood fitness in our experiments, showing its superiority over other generative models in utility tasks. In future work, we aim to further enhance \Algoname's capabilities and applicability in network intrusion detection. This includes integrating reinforcement learning techniques with \Algoname to enable adaptive learning based on real-time threat intelligence, developing algorithms that allow the model to continuously update its parameters in response to new and emerging threats, and conducting extensive field trials to test \Algoname's deployment in various network infrastructures, focusing on scalability, latency, and real-time processing capabilities. This will involve optimizing the model for edge computing environments to ensure low-latency intrusion detection. Additionally, we intend to explore the integration of federated learning with \Algoname, enabling collaborative model training across multiple organizations without the need to share raw data, and developing secure aggregation protocols and differential privacy mechanisms to protect individual data contributions. Beyond network intrusion detection, we will adapt \Algoname to other critical domains such as healthcare and finance by incorporating domain-specific knowledge into the GAN training process, customizing the model architecture and training algorithms to handle the unique data characteristics and requirements of each domain.

\bibliographystyle{IEEEtran}
\bibliography{refs}

\end{document}